# Extended Chaos Theory and Multiparticle Production

*Yi-Fang Chang*
*Department of Physics, Yunnan University, Kunming, 650091, China*
(e-mail: yifangchang1030@hotmail.com)

**Abstract**
First, using the method of the soliton-solution, the fermion probability density equation $d\rho/dE = a\rho(1-2\rho)$, which corresponds to the Dirac equation, is derived. Next, we extend the chaos theory, in which the period bifurcation is equivalent to the particle production. Then this extended chaos theory can be used for description of the multiparticle production and the extensive air showers at high energy. Let the parameter takes a suitable value $\lambda = am_0 = 4 - [A/\ln(E/E_0)]$, the quantitative results will be obtained, and an approximate formula $<n> = a + b\ln S$ will be derived. Many properties of the multiparticle production and of the chaos theory are universal.

## 1. Introduction

Cosmic rays, whose main compositions are some particles with positive charge, for example, proton and alpha-particle, may possess extremely high energies exceeding $10^{17} - 10^{20}$ eV [1,2]. When a charged particle at high energy was converted into the multiparticle production and the extensive air showers (EAS), giant arrays have collecting areas in the range 10-1000 $km^2$ [3,4]. Recently, an important question of high-energy cosmic ray is gamma-ray bursts, and corresponding high-energy neutrinos [5-7]. The usual descriptions of EAS and the multiparticle production at high energies (above $10^{11}$ eV) are mainly various semi-phenomenological models. They are Fermi statistical model, Landau hydrodynamic model, the fireball model, the multiperipheral model, the FF quark cascade model [8] and a class of models based on QCD calculations, including the mini-jet model [9] and Dedenko-Kolomatsky quark-gluon-string model, etc. K.Boruah [10] considered a new mechanism for multiparticle production in the simulation of a cosmic ray cascade. It is the decay of Higgs particle that is produced through vacuum excitation in a cosmic ray collision. A.Lindner [11] reconstructed the energy and determined the composition of cosmic rays in EAS by a new Monte Carlo method.

We proposed [12] that the infinite gravitational collapse of any supermassive stars should pass through an energy scale of the grand unified theory (GUT). After nucleon-decays, the supermassive star will convert nearly all its mass into energy, and produce the radiation of GUT. It may provide some ultrahigh energy sources in astrophysics, for example, quasars and gamma-ray bursts (GRB), etc. This is similar with a process of the Big Bang Universe with a time-reversal evolution in much smaller space scale and mass scale. In this process the star seems be a true white hole. In this paper, we describe quantitatively the multiparticle production and the extensive air showers at high energy by an extended chaos theory.

## 2. From Dirac Field to Extended Chaos Theory

In quantum mechanics, the probability density is
$$\rho = \psi^+ \psi = \overline{\psi}\gamma_4\psi. \qquad (1)$$
For a free Dirac field, the continuity equation is $\partial_\mu j_\mu = 0$, where $j_\mu = \overline{\psi}\gamma_\mu\psi$.

At very high energies the nucleons possess interactions, at least a self-interaction. Based on QCD, the equations should be nonlinear. Since the primary particles of the multiparticle production and of EAS are usually fermions (for example, in cosmic rays), they correspond to the Dirac equation. We suppose that the equation is the Heisenberg unified equation [13]:

$$\gamma_\mu \partial_\mu \psi - b\psi(\psi^+ \psi) = 0, \qquad (2)$$

or a nonlinear Dirac equation

$$\gamma_\mu \partial_\mu \psi + m\psi - b\psi(\psi^+ \psi) = 0. \qquad (3)$$

Eqs.(2) and (3) may be the equations of self-interactions of nucleons (baryons). In a momentum representation, $x_\mu \to p_\mu$,

$$\partial \rho / \partial p_\mu = (\partial \psi^+ / \partial p_\mu)\psi + \psi^+ (\partial \psi / \partial p_\mu) = \gamma_\mu b\rho(2\rho - 1), \qquad (4)$$

where $\psi^+ \psi = 1 - \psi\psi^+$. By using the method of the soliton-solution, let $\eta = \alpha(\gamma_\alpha p_\alpha - u\gamma_4 E)$, Eq.(4) becomes an ordinary differential equation

$$d\rho / d\eta = b\rho(2\rho - 1). \qquad (5)$$

At a mass surface $\vec{p} = 0, a = b\alpha u$, so this equation may be simplified to [14,15]

$$d\rho / dE = a\rho(1 - 2\rho). \qquad (6)$$

For a Dirac field with nonlinear interaction, the continuity equation is

$$\partial j_\mu / \partial x_\mu = \partial \rho / \partial t + \nabla j = a'\rho(1 - 2\rho). \qquad (7)$$

For the momentum representation and $\nabla j = 0$, it is also $d\rho / dE = a\rho(1 - 2\rho)$.

According to Eq.(6), the dimension of a factor $a$ must be [$1/E$]. Let $a = \lambda / m_0$ and $2\rho = X$, the dimension is removed, Eq.(6) turns into a difference equation

$$X_{n+1} = \lambda X_n (1 - X_n). \qquad (8)$$

It is just a logical model, whose chaos solution is well known.

The bifurcation-chaos theory is similar to the multiparticle production and EAS at very high energies in many aspects. In the momentum representation, $X$ becomes unstable and branches continuously with change of a parameter $\lambda$, which corresponds to the energy here. We extend the chaos theory, in which the period bifurcations correspond to the bifurcations of the probability density in Eq.(8), and are equal to the unceasing production of particles. The period corresponds to the multiplicity, which increases as energy. Such a formal resemblance may turn into the quantitative descriptions. In quantum field theory, the total number operator is:

$$N = \sum_i a_i^+ a_i = \int \psi^+(x,t)\psi(x,t)d^3x, \qquad (9)$$

which is similar to the total probability of single-particle theory. The normalized probability $W = \int_{+\infty}^{\infty} \psi^+ \psi dV = 1$ corresponds to the range of $X(0 \le X \le 1)$. When $\lambda \le \lambda_1$, there is a fixed point, which corresponds to a stable particle. When energy is higher than $10^{11}$ eV, the secondary particles are produced from several to tens. When $\lambda$ is equal to or approach to $\lambda_\infty$, the infinite bifurcations correspond to the multiparticle production, the cascade shower and EAS, etc.

**3. Application of Extended Chaos Theory**

Based on the above equations and some experimental results, we make some quantitative researches. Assume that the parameter

$$\lambda = am_0 = 4 - [A / \ln(E / E_0)], \qquad (10)$$

where $A$ and $E_0$ are constants. It is namely that a self-interaction coefficient

$$b = [4 - A / \ln(E / E_0)] / \alpha u m_0$$

is a function of energy in Eq.(3). Here $A / \ln(E / E_0)$ is consistent completely with a coupling constant $g^2(Q^2) = 1 / [a \ln(Q^2 / \Lambda^2)]$ in QCD. Both tend to zero at high energy. They exhibit a characteristic of asymptotic freedom, which has been proved by many experiments. Here $\lambda \in (0,4)$ corresponds to $E \in (E_0'e^{A/4}, \infty)$. Let an input value $E_1 = m_n$ =939.58MeV, then we obtain a table:

| | $E_0'(input)$ | $E_1$ | $E_2(eV)$ | $E_3(eV)$ | $E_4(eV)$ | $E_\infty(eV)$ | $E_0(KeV)$ |
|---|---|---|---|---|---|---|---|
| | $\lambda=0$ | $\lambda_1=3$ | $\lambda_2=3.44949$ | $\lambda_3=3.54428$ | $\lambda_4=3.56446$ | $\lambda_\infty=3.569945$ | |
| a | $10m_e$ | $m_n$ | $2.74298\times10^{11}$ | $3.79341\times10^{12}$ | $7.69145\times10^{12}$ | $9.42813\times10^{12}$ | 898.643 |
| b | $2m_e$ | $m_n$ | $1.58185\times10^{12}$ | $4.92150\times10^{13}$ | $1.24116\times10^{14}$ | $1.62008\times10^{14}$ | 105.106 |

Here $A = \ln(E_1/E_0)$ =6.9523 or 9.0982.

The parameter $\lambda$ may be various different forms, and the input values may also be different, therefore, the values of various bifurcation points are also changeable. For the electromagnetic cascade showers, let $E_1 = m(\pi^0) = 135$ MeV, then we may obtain a similar table.

Eq.(9) is replaced into Eq.(6), the equation becomes
$$d\rho/dE = m_0^{-1}[4 - A/\ln(E/E_0)]\rho(1-2\rho), \qquad (11)$$
where $m_0$ is a constant whose dimension is energy. According to the physical meanings of various quantities, we may suppose $E/E_0 = D\sqrt{S}, <n> = c\rho$, then Eq.(11) becomes a Bernoulli equation
$$\frac{d<n>}{d\sqrt{S}} = \frac{DE_0}{m_0}(4 - \frac{A}{\ln D\sqrt{S}})(<n> - \frac{2<n>^2}{c}). \qquad (12)$$
For the first approximation, we solve Eq.(12), and derive a solution:
$$<n> \approx (1 + A\ln D/4)(c/2m_0) + (Ac/16m_0)\ln S = a + b\ln S. \qquad (13)$$
Applied an approximation
$$\int \exp[-\frac{(A+4)E}{2}](\frac{E}{E_0})^{2E}\ln(\frac{E}{E_0})dE \approx \frac{2}{8-A}e^{-(A+4)E/2}(\frac{E}{E_0})^{2E}\ln\frac{E}{E_0}$$
at high energies, then
$$<n> \approx \frac{c}{2m_0}(1 - \frac{8}{A})[(1 + \frac{4}{A}\ln D) + \frac{2}{A}\ln S + \frac{4}{A^2}(\ln S)^2]. \qquad (14)$$

These formulas are consistent with the prediction of the average particle number is $<n> \sim \ln S$ by the multiperipheral model, the Mueller-Regge analysis and the bremsstrahlung model [16], etc. For the *pp* interaction with energy scale (3-152GeV) [17],
$$<n_{ch}> = (1.17 \pm 0.10) + (0.30 \pm 0.05)\ln S + (0.13 \pm 0.01)(\ln S)^2.$$

They are the relations between a changed process of single-particle, or the collision process of particle-cluster and their energies. Combing other models, for example, the fireball model, it is namely a changed process of one fireball, two or many fireballs.

Using an equation
$$\partial\rho/\partial x_\mu = a\rho(1-2\rho), \qquad (15)$$
in the space-time, the dimension of parameter $a^{-1}$ is the same with time (life) or space (distance). Therefore, the equation will be able to describe similarly the relations between the multiplicity and time (life) or distance in the multiparticle production when the energy was fixed.

The universality of the chaos theory and the general constants $\delta, \alpha$ ,etc., are independent of the specific forms of the transformations $f(x)$ (i.e., the equations). It may explain just that the general scaling of the multiparticle production is independent of the particle-types and of the interacting ways at high energy; and explain that a law on the average multiplicity <n> is the same for various hadrons. Further, some quantitative relations on the multiparticle production should be independent of the specific forms of equations and nonlinear interactions. The self-similarity property of the multiparticle production process [18,19] shows a fractal characteristic, which may be derived from the chaos theory. Cao and Hwa [20,21] have investigated the chaoticity of multiparticle production in high-energy collision. The development of shower is a stochastic process, while chaos is also an internal stochasticity. Therefore, many properties and corresponding methods should possess some similarities among the multiparticle production, EAS, chaos and the extended chaos theory.